# Magnetization Reversal by Electric-Field Decoupling of Magnetic and Ferroelectric Domains Walls in Multiferroic-Based Heterostructures

V. Skumryev,[1, 2] V. Laukhin,[1, 3] I. Fina,[3] X. Martí,[3] F. Sánchez,[3] M. Gospodinov,[4] and J. Fontcuberta[3*]

[1]*Institució Catalana de Recerca i Estudis Avançats (ICREA), Catalonia (Spain).*
[2]*Departament de Física, Universitat Autònoma de Barcelona, 08193 Bellaterra, Catalonia (Spain),*
[3]*Institut de Ciència de Materials de Barcelona ICMAB-CSIC, Campus UAB, Bellaterra, Catalonia (Spain),*
[4]*Institute of Solid State Physics, Bulgarian Academy of Sciences, 1784 Sofia, Bulgaria*

ABSTRACT

We demonstrate that the magnetization of a ferromagnet in contact with an antiferromagnetic multiferroic ($LuMnO_3$) can be speedily reversed by electric field pulsing, and the sign of the magnetic exchange bias can switch and recover isothermally. As $LuMnO_3$ is not ferroelastic, our data conclusively show that this switching is not mediated by strain effects but is a unique electric-field driven decoupling of the ferroelectric and ferromagnetic domains walls. Their distinct dynamics are essential for the observed magnetic switching.

With their rich physics, multiferroics (materials in which magnetic and polar orders coexist) have emerged as some of the most promising materials for multifunctional applications in spintronics, owing to the advantageous possibility of controlling the magnetic state by electric fields and vice versa [1-6]. However, in spite of expectations, the minute magnetoelectric coupling of both order parameters [7] in single-phase multiferroics has hampered device-driven progress and focus is being directed towards exploring interface coupling either *via* strain [8,9] or by the exchange interaction between an AF multiferroic and a ferromagnet (FM) giving rise to magnetic exchange bias (EB). The best known EB manifestation is the magnetic hysteresis loop-shift along the field axis, when the system is cooled trough the Néel temperature in magnetic field. This shift can be either in the "negative" [10] or in the "positive" field direction [11] and it is largely employed in spin valves and magnetic tunnel junctions.

The EB in magnetoelectrics was first explored [12] at the interface of the archetypical magnetoelectric (without long range polar order) $Cr_2O_3$ with a FM layer and a sign switch of the EB was observed depending on the electric field applied during the cooling procedure trough the Néel temperature ($T_N$) of $Cr_2O_3$. Following the original observation of coupled order parameters in hexagonal manganites [13], a substantial electric-field-induced suppression of the EB was found using the multiferroic $YMnO_3$ [6], shortly after EB in $YMnO_3$ had been demonstrated [14,15]. Those pioneering works were soon followed by studies [16-20] on $BiFeO_3$, which so far is the only room-temperature multiferroic. Although it has been proposed [16-18] that EB could be related to the ferroelectric and AF domain structure of $BiFeO_3$, which can be modified by an electrical field [17-19, 21], strain-mediated coupling has not been excluded [22]. However, to date it has not been demonstrated that the sign of the EB can be reversibly switched at a certain temperature by electric field, without the need to follow the usual field-cooling protocol, which is incompatible with the most obvious applications in spintronics. In this Letter we show that the sign of EB can be switched and reset by appropriate electric pulses, without the need of varying setting temperature. It will be shown here that electric-field driven decoupling of the ferroelectric and ferromagnetic domains walls, and their distinct dynamics, are essential for the observed magnetic switching.

We demonstrate this EB resetting for the magnetization of 10 nm thin ferromagnetic $Ni_{81}Fe_{19}$ (Py) film deposited on the basal plane of hexagonal $LuMnO_3$ single crystal (~12 um thick). Py is a soft FM material, with coercivity of only few Oersteds, while $LuMnO_3$ is an AF multiferroic isostructural to $YMnO_3$ [23], with coupled antiferromagnetic ($T_N$ = 93 K) [24] and polar ($T_C$ > 570 K) [25] order parameters. It has a hexagonal crystal structure [23] and below $T_N$ the Mn moments adopt a triangular arrangement within the basal plane.

Cooling down the sample trough $T_N$ in magnetic field of 3 kOe applied in the basal plane, resulted in establishing a positive EB field, $H_{eb}$ = + 130 Oe as evidenced by the negative hysteresis loop shift shown in Fig 1a (orange arrow). All measurements were performed at 5 K, cycling H between +3 and -3 kOe. Training effect is common in exchange biased systems. Here it is reflected by the difference between the virgin curve (labelled "$v$") and the 1st consecutive loop (labelled "$t_1$") in Fig.1a and it quickly dies off for further cycles. Hence, in our experiments to minimize training effects, we choose to apply electric field to the 2nd trained curves ("$t_2$"); however, the results reported in the following are found to be qualitatively similar for any set of magnetization loops.

Applying electric pulse of 40 V ($\approx$ 3 $10^6$ V/m), duration 500 μs and triangular profile, at the descending branch of the magnetization loop, at H = –145 Oe (a value chosen to be close to the coercive field), causes an abrupt jump of the Py magnetic moment (① → ② in Figs. 1a and 1b), which also changes its direction. Up to this point we qualitatively reproduced the result already reported using $YMnO_3$ film [6]. Further cycling of magnetic field after the application of this electric pulse (curves $v$, $t_1$ and $t_2$ in Fig. 1b), results in concessive hysteresis loops being close to the *mirror image* of those before the pulse shown in Fig. 1a, and importantly, the corresponding $H_{eb}$ has its sign opposite ($H_{eb}$ = -55 Oe, $v$-curve, orange arrow) to the original one. It is relevant to note, that thought qualitatively similar, the two sets of loops - before and after the electric pulse - have some evident differences: $H_{eb}$ is smaller (in modulus) and the coercive field $H_C$ is larger after the pulse (Supplementary Information). At this point it was tempting to see if at the settled temperature an electric pulse could recover the polarity of the original EB. Indeed, application of the electric pulse of 40 V at the ascending branch of the magnetization loop, at H = +60 Oe, triggers magnetization reversal (③ → ④, in Figs. 1b and 1c). Subsequent field cycling leads to hysteresis loops (Fig. 1c) very similar to those shown in Fig. 1a (recorded after the initial field-cooling process trough $T_N$). Importantly, comparison to data in Fig 1a, indicates that the sign of the EB field ($H_{eb}$ = +110 Oe, *v*-loop, orange arrow) has been recovered.

Similar electric-induced switching of magnetization was observed for other pulse amplitudes. For example, after using the same experimental protocol, a pulse of 5 V triggered an abrupt jump of the Py magnetic moment to a value very similar to one measured after the 40 V pulse. The subsequent EB ($H_{eb}$ = -22 Oe) is, however smaller compared to the one after applying 40 V. Comparison of the EB after electric pulses, in the *t*-loops indicates the same trend, as shown in Fig. 2a where we collect the dependence of the $H_{eb}$ (minus the loop shift) on the applied voltage for *v*- and *t*-loops.

Fig. 2a illustrates that the EB strongly depends, particularly for low voltages, on the applied voltage. This indicates that the underlying mechanism of EB switching weakens with decreasing the electrical field amplitude: whereas it was repeatedly observed down to 5 V, no switching was observed below a threshold of about 2 V. Switching was found to be insensitive to pulse polarity.

The observed peculiar EB switching and resetting, triggered by an electric field has no analogue in the conventional exchange bias systems. It requires that the FM moments of the Py become unpinned by electrical pulse, but then pinned again in direction opposite to the initial one. This let us to suggest the polar order of the AF LuMnO$_3$, namely its ferroelectric domains walls, as a prime suspect behind the observed EB reset.

Due to strong uniaxial anisotropy of the hexagonal LuMnO$_3$, only two kinds of 180° ferroelectric (FE) domains with opposite sign of the FE order parameter are expected to exist, with narrow domain walls (FE-DWs) between them that span only few unit cells [26]. On the other hand, 180° AF domains [27] also exist. Their walls carry [27,28] net magnetic moment. EB at compensated magnetic surfaces, such as (0001) of LuMnO$_3$, owes its existence exclusively to these uncompensated moments. In LuMnO$_3$, the small basal plane anisotropy of the Mn moments should render AF-DWs at least two orders of magnitude wider ($10^2$-$10^4$ unit cells) than the FE-DWs, and their net moment should be in the basal plane [26]. In hexagonal RMnO$_3$ (R=Y, Lu, etc) it has been shown that FE and AF domains are clamped and two distinct types of AF-DWs exist [13,29]: 'clamped' walls (c-AF-DW) formed at the position of all FE-DWs, and "unclamped" walls (u-AF-DW) formed within the large ferroelectric domains . c-AF-DWs are created and clamped at the centre of each of the already existing FE-DWs to minimize the free energy [26].

The scenario, which we propose for explaining the observed peculiar effect of the electric pulses on the EB, begins with cooling down the sample in a presence of magnetic field to ensure that the Py moments are aligned. Below $T_N$, clamped and unclamped AF-DWs are formed. The orientation of the uncompensated moment of the AF-DWs is dictated by the exchange interactions with the ferromagnetic Py moments at the interface. The uncompensated moments in both types of AF-DWs serve as pining sides for the Py moments giving rise to EB as illustrated in the insets of Fig 1. Once c-AF-DWs are formed at FE-DWs, it is difficult to separate them from the FE-DWs since an energy barrier has to be overcome [26]. The magnetic hysteresis cycle after the cooling procedure (Fig.1a) should be qualitatively similar to the one for conventional EB systems; the only difference being that two different types of AF-DWs contribute to pinning. Whereas u-AF-DWs do not correlate with FE domains and thus should always provide EB unaffected by the multiferroicity, this is not the case for the c-AF-DWs which contribution to EB should be affected by FE-DW motion under electric field.

Suggesting the way, in which the electric field assists the magnetization rotation, is obviously crucial for elucidating the observed EB switching and subsequent reset. Since no irreversibility is found on the dielectric response of LuMnO$_3$ up to 40 V (see the Supplementary Information) switching of overall polarization can be ruled out as a driving force. The experimental results from Fig. 1 clearly indicate that under application of the electric pulse, some AF-DWs have been unpinned – at least for a short period of time - with the concomitant collapse of the pinning chain providing the EB, which is cancelled and the DW-magnetic moments can freely rotate under the external magnetic field; subsequent pinning of the AF-DWs should lead to a new EB, eventually of sign opposite to the initial one. Obviously, the c-AF-DWs are the candidates to play such key role.

We propose that applying a pulse (① in Figs. 1) with a suitable amplitude and duration could quickly expand some FE domains and thus move the FE-DW far from its initial position (Fig 3a) leaving behind unpinned the c-AF-DW that was sitting at its centre (Fig. 3b). This reduces the AF pinning force on the Py moments and allows magnetic field to reverse them instantly, forced by the (negative) applied magnetic field at the time of the electrical pulse. After the electric pulse, AF-DW should be re-established at a FE wall (which could be in a new position or back at the initial one, depending whether the movement was reversible or irreversible) to minimize the free energy. Importantly, the magnetic moment of the newly pinned c-AF-DW will be in a direction opposite to the one before the electrical pulse since its orientation will be dictated by the reversed magnetization of Py (point ① → ② in Fig 1), which reaches about 90 % of its saturation value, as well as the polarity of the applied field. This will impose a pinning direction opposite to that before the electrical pulse, as observed in Figs. 1b and sketched in Fig 3c.

When the second electrical pulse is applied at point ③ of Fig. 1 (Fig. 3d), the EB will be reset, (Fig. 3e): i.e. the initial EB set trough field cooling procedure trough $T_N$ will be recovered. Note that, in this scenario and in agreement with experimental data, the changes of magnetization and EB are independent of the voltage polarity. However, within this picture, a threshold V should exist for unpinning the c–AF-DW from FE-DW centre. Indeed, if the displacement of the FE-DW is not large enough, the magnetic moments of the AF-DW will not be unpinned. This is in agreement with the results from Fig. 2a, which reveal that electrical pulse of 2 V is incapable to switch magnetization. Similarly, one should expect that if the speed at which the FE-DW travels under the electric field stimulus is not fast enough, the AF wall would be able to keep "clamped" to it during the excursion and no effect will be observed. To verify this prediction we have performed experiments using V(t) pulses of distinct duration (rising-falling) time. In Figs. 2b and 2c we show the switching (③ → ④') induced by a pulse of V = 40 V with a duration of 500 ms and 100 s, respectively. It is clear that the magnetization is only partly switched for longer (smaller $dV/dt$) pulses. In a subsequent experiment, if the slower pulse (100 s, Fig. 2c) is followed by another 40 V pulse with duration 500 μs (④' →④) the magnetization jump is further enhanced until the saturated final state is recovered.

The two types of AF domain walls, the "clamped" and the "unclamped" ones play quite distinct roles in the proposed scenario. While the former could alternatively change the sign of the unidirectional anisotropy imposed on the Py moments after each concessive electric pulse, the later always keep the initially established direction. Thus while the "unclamped" AF walls will always reinforce the initial EB established via field cooling procedure trough $T_N$ (light-gray arrows in Fig 1), the "clamped" walls could alternatively produce either positive or negative EB after each concessive electric pulse (dark-gray arrows in Fig 1). The difference in the hysteresis loops prior and after application of electrical pulse should also depend on the number of "clamped" AF-DWs that are left behind the FE-DW excursion and thus on the pulse amplitude as actually observed in our experiments and summarized in Fig. 2a. It is worth noting that situation with domain walls pinned in two opposite directions could lead to smaller loop shift and thus an overall decrease in EB, although in agreement with experimental observation (Fig. 1) $H_C$ could be increased.

Finally, observation of interface-mediate exchange coupling in $LuMnO_3$ is relevant because, in this case and in contrast to $BiFeO_3$ [22], strain is not expected to play a role in domain coupling and switching as $LuMnO_3$ is not ferroelastic.

In conclusion, we have shown that at a settled temperature, the sign of EB can be switched by an electric pulse and reset after applying certain magnetic field followed by a second electric pulse. This phenomenon has no analogy in the conventional EB based on antiferromagnetic but non-multiferroic materials. It constitutes a clear evidence of reversible control of magnetization using electric field without varying the settled temperature. The scenario proposed to explain our findings emphasizes the distinct role played by the AF domain walls: "clamped" *vs* "unclamped" to the FE ones. The "clamped" AF-DW´s appear to be responsible for the electrical tunability of the pinning torque exerted on the FM moments. Since the "unclamped" AF-DWs are often born on structural defects, while the number of "clamped" is predetermined by the number of FE domains, one should be able to tune the hysteresis loops and the exchange bias by controlling the ratio between them (for example introducing structural defects by irradiation or varying the number of FE domains by the thickness of the multiferroic).

The proposed scenario results from the dynamics of the AF and FE domain walls rather than from their density, which itself may determine the absolute values of EB as found in BiFeO$_3$ [16,19]. It is based on the assumption that the AF and the FE domain walls can not respond in *tandem* in response to the external stimulus, thus implying distinct dynamic properties and effective masses. It requires the ability to push FE domain walls away from the AFM ones; owing the dimensions of the later and typical FE domain wall motion velocities, responses down to the picoseconds regime could be achieved. Further studies on coupled FE and AF domain-wall dynamics are needed to get a more detailed microscopic understanding. In any event, the possibility of ultrafast switching and modulating the exchange bias at fixed temperature by electric field open new possibilities, particularly if the same phenomena could be identified in room-temperature multiferroics.

*Note add in proof*.— When this manuscript was completed, we learned of two very recent Refs [31,32] reporting isothermal EB sign switching using Cr$_2$O$_3$ and BiFeO$_3$, respectively. However, we note that the suggested mechanisms behind the claimed effects are completely different from the one proposed in our work.

Acknowledgements: Financial support by the Ministerio de Ciencia e Innovación of the Spanish Government [Projects MAT2008-06761-C03 and NANOSELECT CSD2007-00041] and Generalitat de Catalunya (2009 SGR 00376) is acknowledged.

**References**

[1] M. Bibes and A. Barthélémy, Nature Mater. **7**, 425 (2008).

[2] S.W. Cheong and M. Mostovoy, Nature Mater. **6**, 13 (2007).

[3] R. Ramesh and N.A. Spaldin, Nature Mater. **6**, 13 (2007).

[4] M. Gajek *et al.*, Nature Mater. **6**, 296 (2007).

[5] H, Zheng *et al.*, Science **303,** 661 (2004).

[6] V. Laukhin *et al.*, Phys. Rev. Lett. **97**, 227201 (2006).

[7] W.H. Meiklejohn and C.P. Bean, Phys. Rev. **102**, 1413 (1956).

[8] W.H. Meiklejohn and R.E. Carter, J. Appl. Phys. **30**, 2020 (1959).

[9] W. Eerenstein, N.D. Mathur, and J.F. Scott, Nature **442**, 759 (2006).

[10] W. Eerenstein *et. al.*, Nature Mater. **6**, 348 (2007).

[11] S. Geprägs *et al.*, Appl. Phys. Lett. **96**, 142509 (2010).

[12] P. Borisov *et al.*, Phys. Rev. Lett. **94**, 117203 (2005).

[13] M. Fiebig *et al.*, Nature **419**, 818 (2002).

[14] J. Dho and M.G. Blamire, Appl. Phys. Lett. **87**, 252504 (2005).

[15] X. Martí et al., Appl. Phys. Lett. **89**, 032510 (2006).

[16] H. Bea *et al.,* Phys. Rev. Lett. **100**, 017204 (2008).

[17] Y.H. Chu *et al.,* Nature Mater. **7**, 478 (2008).

[18] D. Lebeugle *et al.*, Phys. Rev. Lett. **103**, 257601 (2009).

[19] L.W. Martin *et al.,* Nano Lett. **8**, 2050 (2008).

[20] J. Dho and M.G. Blamire, J. Appl. Phys. **106**, 073914 (2009).

[21] S. Lee *et al.*, Appl. Phys. Lett. **19**, 192906 (2008).

[22] N.D. Mathur, Nature **454**, 591 (2008).

[23] B.B. Van Aken *et al.*, Nature Mater. **3**, 164 (2004).

[2]. V. Skumryev *et al.*, Phys. Rev. B **79**, 212414 (2009).

[25] E.F. Bertaut. *et al.*, Acad. Sci. Paris **256**, 1958 (1963).

[26] A.V. Goltsev *et al.*, Phys Rev Lett. **90** 177204 (2003).

[27] L. Néel, Proc. of the Int.Conf. on Theoretical Physics, Kyoto, September, 1953, Science Council of Japan **701** (1954).

[28] Y.Y. Li, Phys. Rev. **101** 1450 (1956).

[29] Th. Lottermoser and M. Fiebig, Phys. Rev. B **70**, 220407 (2004).

[30] F. Yen *et al.*, *J. Mater. Res.* **22**, 2163 (2007).

[31] Xi He *et al.*, Nature Materials **9**, 579 (2010)

[32] S. M. Wu *et al.*, Nature Materials **9**, 756 (2010)

**Figures Caption**

**Figure 1.** Concessive hysteresis loops after field cooling procedure and cartoon of the experimental set up. In (a) magnetization loops recorded after cooling the sample under 3 kOe field from well above $T_N$ (virgin loop (*v*) – circles) and trained loops recorded by subsequent isothermal field cycling ($t_1$ and $t_2$) – triangles (solid and empty, respectively). The red dashed arrow indicates the magnetic moment jump from point r ① where electric pulse is applied to point ②. (b) virgin (*v*) and trained (*t*) loops after 40 V pulse at point ① in panel a; (c) loop after applying the 40 V pulse at point ③. Legend for the insets is shown in the top right area of the Figure.

**Figure 2.** Dependence of the magnetic hysteresis loop shift and magnetic moment modifications on the characteristics of electric pulses. (a) Magnetic hysteresis loop shift (- $H_{eb}$) (*v*-loops and *t*-loops) after the electric pulse as a function of the electric pulse amplitude; (b) Magnetization changes after applying 40 V pulse with duration 500 ms; (c) after applying 40 V pulse with duration 100 s, followed by 40 V pulse with duration 500 us

**Figure 3.** A sketch (top view of the basal plane) of the position of "clamped" FE and AF domain walls pair at different points of the hysteresis loops depicted on Figures 1 and 2. From left to right: at point ① ; during the magnetization jump from ① to ② caused by the 1$^{st}$ electric pulse, when "unclamping" takes place; at point ② where a "clamping" is re-established but the direction of the AF domain wall moment is reversed

**Figure 1**

**Figure 2**

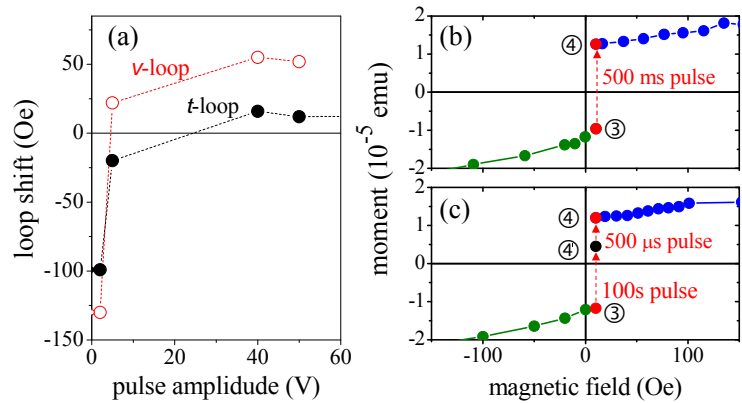

**Figure 3**

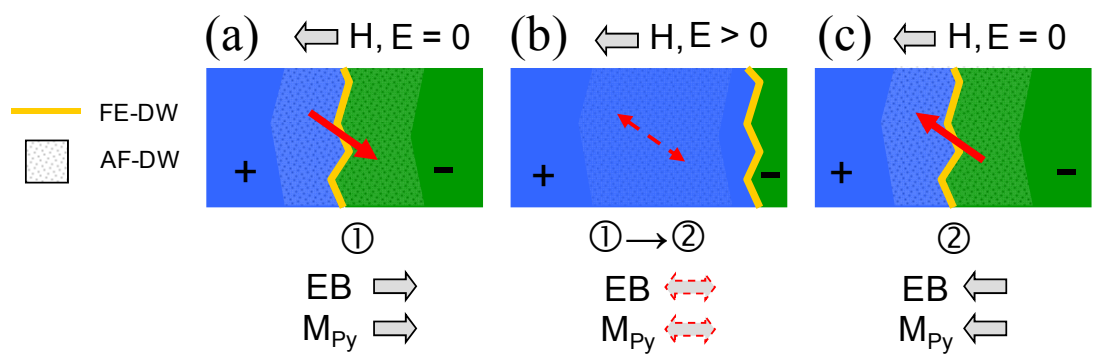